\documentclass[11pt]{article}

\usepackage[margin=1in]{geometry}
\usepackage[T1]{fontenc}
\usepackage[utf8]{inputenc}
\usepackage{lmodern}
\usepackage{microtype}
\usepackage[numbers,sort&compress]{natbib}
\usepackage{amsmath,amssymb}
\usepackage{booktabs}
\usepackage{array}
\usepackage{enumitem}
\usepackage{hyperref}
\usepackage{xcolor}
\setlist[itemize]{leftmargin=1.5em,topsep=3pt,itemsep=2pt,parsep=0pt}
\setlist[enumerate]{leftmargin=1.5em,topsep=3pt,itemsep=2pt,parsep=0pt}
\hypersetup{
  pdftitle={Execution Envelopes: A Shared Admission Contract for Backend AI Execution Requests},
  pdfauthor={Krti Tallam},
  pdfsubject={Normalized backend admission contracts for AI execution requests},
  pdfkeywords={admission control, execution envelope, ai infrastructure, authorization, scheduling, serving},
  colorlinks=true,
  linkcolor=blue!50!black,
  citecolor=blue!50!black,
  urlcolor=blue!50!black
}

\title{\textbf{Execution Envelopes: A Shared Admission Contract for Backend AI Execution Requests}}
\author{Krti Tallam\\
Kamiwaza AI, San Francisco, CA, USA\\
\texttt{krti@kamiwaza.ai}}
\date{May 2026}

\begin{document}
\maketitle

\begin{abstract}
Enterprise AI backends increasingly admit heterogeneous execution requests across model deployment, inference, evaluation, data movement, and agentic workflows. In many systems, those requests arrive in service-specific shapes, which makes it difficult to attach shared admission-time behavior such as logging, governance hints, resource accounting, authorization-aware policy hooks, and later runtime review without rebuilding the same contract in each subsystem. This paper introduces the \emph{execution envelope}, a normalized internal admission object that records who is asking for what kind of execution, what resources were requested, what policy-relevant scope accompanied the request, and what the backend ultimately granted. The proposal is intentionally narrow. It does not replace service-specific request models, perform scheduling, or introduce a new authority token. Instead, it defines a descriptive admission seam that can be threaded through real backend paths before backend-specific resolution begins. I formalize the distinction between requested and granted resources, specify the field families, invariants, and lifecycle of the envelope, work through \texttt{POST /serving/deploy\_model} as an initial proving ground, and position the design relative to usage control, analyzable authorization, admission control, and cluster scheduling. The central claim is that a shared execution-admission contract is a useful missing primitive for modern AI backends because it creates one place to attach governance and observability without pretending to solve placement, policy, and runtime execution in a single step.
\end{abstract}


\section{Introduction}

Modern AI platforms do not handle one kind of execution. They handle model deployment, online inference, data preparation, evaluation jobs, workflow steps, agent runtime actions, and internal maintenance tasks. The systems challenge is not only that each class of work has different resource needs. It is also that admission semantics drift across services. Identity may be normalized in one place, resource ask fields may be rich in another, governance hints may exist only in logs, and backend-resolved allocations may not be recorded in a common shape at all.

This fragmentation matters because shared backend concerns increasingly sit at admission time. Operators want a stable place to attach request logging, guardrail hooks, resource accounting, policy-driven model governance, and eventually richer forms of execution control. But if every backend path encodes ``who asked for what'' differently, then each new concern has to reconstruct the same contract repeatedly.

Why this matters now is that agentic platforms are no longer composed only of static deployment and inference endpoints. They increasingly mix delegated workflows, retrieval jobs, model lifecycle actions, and runtime tool execution under one operational control plane. Once those paths share identity, tenancy, and governance surfaces, the lack of a shared execution-admission description becomes more than a code-quality annoyance. It becomes a platform-level blind spot. The backend may know who the requester was, and it may eventually know what runtime target was selected, while still lacking one durable object that says what execution was originally being admitted and how that request changed as backend resolution proceeded.

This paper proposes a narrow systems primitive for that gap: the \emph{execution envelope}. The envelope is an internal admission object that records:

\begin{itemize}
    \item who asked;
    \item what execution was requested;
    \item what scope or governance hints accompanied the request;
    \item what resources were requested;
    \item what resources and routing details the backend ultimately granted.
\end{itemize}

The proposal is intentionally constrained. It is not a scheduler, not a placement engine, not a new public API payload, and not a replacement for authorization. It is a shared descriptive contract built early enough in the request path that later systems do not need to reverse engineer intent and context after the fact.

The contribution of the paper is fourfold. First, it argues that current AI backends have a missing admission primitive distinct from both authorization and scheduling. Second, it specifies a concrete envelope model, invariants, and lifecycle. Third, it defends a deployment strategy in which the envelope is proven on one path, \texttt{POST /serving/deploy\_model}, before wider rollout. Fourth, it positions the design relative to adjacent work in usage control, analyzable authorization, admission control, and large-scale cluster scheduling.

\section{Problem Statement}

Most mature backends already have some notion of request context. They know identity, tenant, request ID, and perhaps the service entrypoint. What they often lack is a normalized description of the requested execution itself.

The problem can be stated simply:

\begin{quote}
backend identity is often normalized earlier than backend execution.
\end{quote}

This asymmetry creates several operational issues.

\subsection{Repeated reconstruction}

When every service-specific path encodes execution shape differently, each new concern---logging, governance, admission policy, or accounting---must reconstruct the same request semantics locally.

\subsection{Loss of original intent}

Many systems preserve the final resource decision but not the caller's original ask, or vice versa. That makes later reasoning about oversubscription, denial, or narrowing much harder.

\subsection{Premature coupling}

Without a narrow shared contract, teams often jump too quickly from ``we need consistent admission semantics'' to ``we need a new scheduler'' or ``we need a new authorization model.'' Those are larger and riskier changes. The execution envelope is intended to separate the admission-description problem from both.

\subsection{Degraded-path opacity}

The failure mode is not only inconsistency. It is opacity under change. Backends regularly narrow requests, bind them to specific runtimes, or reject them after partial resolution. If there is no common admission artifact, those transitions become local implementation facts rather than platform-visible state. Later operators can often recover the final outcome or the original caller, but not the path between them. In contemporary AI backends, that gap matters because many governance questions are about the transition itself: what the caller asked for, what was denied or narrowed, and what the backend ultimately permitted.

\section{Design Goals and Non-Goals}

\subsection{Goals}

The execution envelope is designed to:

\begin{enumerate}
    \item normalize execution-capable requests across human-, app-, and service-initiated paths;
    \item preserve caller intent separately from backend resolution;
    \item provide one internal seam for logging, governance hints, and later downstream consumers;
    \item land in one real path so it becomes testable infrastructure rather than a whiteboard abstraction;
    \item remain additive enough that the first consumer can adopt it without changing public API or placement semantics.
\end{enumerate}

\subsection{Non-goals}

The proposal explicitly does not:

\begin{enumerate}
    \item make scheduling or placement decisions;
    \item replace existing service request models;
    \item change public endpoint payloads in the first slice;
    \item require collaboration workspace scope for every request;
    \item introduce a new authority token or permission calculus.
\end{enumerate}

\section{Execution Envelope}

The execution envelope is an internal backend object with seven field families.

\begin{table}[htbp]
\centering
\small
\caption{Execution envelope field families.}
\label{tab:fields}
\begin{tabular}{>{\raggedright\arraybackslash}p{2.2cm} >{\raggedright\arraybackslash}p{10.6cm}}
\toprule
\textbf{Field family} & \textbf{Role} \\
\midrule
\texttt{caller} & normalized requester identity and request metadata copied from existing request context \\
\texttt{execution} & descriptive classification of what is being requested: kind, source service, operation \\
\texttt{scope} & optional workspace, app, job, session, or thread identifiers when available \\
\texttt{governance} & optional hints for logging, audit, guardrails, sensitivity, or other policy surfaces \\
\texttt{requested resources} & caller intent: cpu, memory, gpu, concurrency, timeout, priority, and similar ask fields \\
\texttt{granted resources} & backend-resolved allocation after service-specific validation and narrowing \\
\texttt{resolution} & runtime metadata such as backend kind, actor binding, routing target, or execution substrate \\
\bottomrule
\end{tabular}
\end{table}

\begin{table}[htbp]
\centering
\small
\caption{Illustrative typed core fields for a minimal envelope artifact.}
\label{tab:typed-fields}
\begin{tabular}{>{\raggedright\arraybackslash}p{4.0cm} >{\raggedright\arraybackslash}p{2.1cm} >{\raggedright\arraybackslash}p{6.7cm}}
\toprule
\textbf{Field} & \textbf{Type} & \textbf{Meaning} \\
\midrule
\texttt{envelope\_id} & string & stable admission identifier used to correlate request and resolution \\
\texttt{caller.requester\_urn} & string & normalized user or service identity \\
\texttt{execution.kind} & enum & deployment, inference, pipeline step, maintenance task, etc. \\
\texttt{execution.operation} & string & service-local operation name such as \texttt{deploy\_model} \\
\texttt{requested.gpu\_count} & integer & original caller ask for GPU quantity \\
\texttt{requested.engine} & string & requested backend or engine family \\
\texttt{granted.gpu\_count} & integer & validated or narrowed GPU allocation \\
\texttt{resolution.backend} & string & resolved execution substrate or backend class \\
\bottomrule
\end{tabular}
\end{table}

\subsection{Requested versus granted}

The strongest semantic choice in the design is the distinction between \texttt{resources\_requested} and \texttt{resources\_granted}. The envelope should preserve original caller intent even when the backend narrows, rejects, or rebinds the ask. Overwriting the request with the grant destroys useful information.

Let $r_q$ denote the resource vector requested by the caller and $r_g$ the resource vector ultimately granted by the backend. The envelope's role is not to compute the mapping
\begin{equation}
r_q \mapsto r_g
\end{equation}
in the abstract. The role is to preserve both values under the same request identity so that admission-time and resolution-time reasoning remain joinable.

\subsection{Descriptive, not decisive}

The envelope remains descriptive at admission time. Backend-specific components retain authority over validation and resolution. This keeps the shared contract honest: it does not pretend the API layer can allocate resources or determine placement for a heterogeneous fleet.

\subsection{Contract invariants}

The envelope becomes much more useful if a few invariants are explicit rather than implied.

\begin{enumerate}
    \item \textbf{Admission immutability.} After construction, \texttt{caller}, \texttt{execution}, \texttt{scope}, \texttt{governance}, and requested resources should be immutable except for additive annotation. Otherwise later consumers cannot trust the original ask.
    \item \textbf{Grant append-only semantics.} Backend components may populate or refine granted resources and \texttt{resolution}, but they should not rewrite the original request fields in place.
    \item \textbf{Stable correlation.} One request identifier should join the admitted envelope, backend resolution, and subsequent logging so that accounting and audit are reconstructable.
    \item \textbf{Namespaced extension.} Service-specific fields should attach in explicitly namespaced extension blocks rather than overloading the shared core contract.
\end{enumerate}

\section{Worked Example: \texttt{deploy\_model}}

The paper's narrowness is easiest to evaluate against one real path. Table~\ref{tab:deploy-model-example} shows how a model-deployment request maps into the envelope.

\begin{table}[htbp]
\centering
\small
\caption{Illustrative envelope population for \texttt{POST /serving/deploy\_model}.}
\label{tab:deploy-model-example}
\begin{tabular}{>{\raggedright\arraybackslash}p{2.5cm} >{\raggedright\arraybackslash}p{10.2cm}}
\toprule
\textbf{Envelope family} & \textbf{Example value or interpretation} \\
\midrule
\texttt{caller} & authenticated user or service principal, tenant, request ID, timestamp \\
\texttt{execution} & \texttt{kind=deployment}, \texttt{service=serving}, \texttt{operation=deploy\_model} \\
\texttt{scope} & optional workspace, app, template, or deployment-group identifier if present \\
\texttt{governance} & audit level, sensitivity tag, deployment provenance, or guardrail hint if supplied \\
requested resources & desired engine, gpu count, cpu/memory ask, placement preference, timeout, concurrency hints \\
granted resources & validated engine choice, narrowed resource limits, selected backend class, or actor binding \\
\texttt{resolution} & deployment target, runtime substrate, backend path, or scheduling handoff metadata \\
\bottomrule
\end{tabular}
\end{table}

This example matters because it demonstrates that the envelope is not speculative metadata. It can be built entirely from existing request context and request fields while preserving the distinction between what the caller asked for and what the backend accepted. In the repository's concrete artifact, a tenant administrator requests a \texttt{vllm}-backed deployment of a specific model configuration, asks for one full GPU-equivalent allocation and a particular location preference, and carries an explicit audit event hint. Later phases preserve those requested facts while the backend resolves a narrower granted allocation, runtime target, and deployment identity.

\section{Lifecycle}

The initial lifecycle is deliberately simple, but it is richer than a single admitted-or-not bit. In the first proving-ground implementation, the same execution envelope is persisted across distinct phases rather than collapsed into one opaque event.

\begin{enumerate}
    \item a supported endpoint admits a request and builds an execution envelope from request context and request payload;
    \item the backend service performs its existing validation and resolution logic;
    \item the backend may populate \texttt{resources\_granted} and \texttt{resolution};
    \item the envelope is recorded in internal metadata and logging for later consumers.
\end{enumerate}

This lifecycle matters because it avoids two common design mistakes. First, it avoids pretending that placement can be standardized too early. Second, it keeps identity, requested execution, and resolved execution attached to the same object.

\begin{table}[htbp]
\centering
\small
\caption{Initial \texttt{deploy\_model} lifecycle phases and what each phase adds.}
\label{tab:lifecycle}
\begin{tabular}{>{\raggedright\arraybackslash}p{2.1cm} >{\raggedright\arraybackslash}p{4.5cm} >{\raggedright\arraybackslash}p{5.8cm}}
\toprule
\textbf{Phase} & \textbf{What is preserved} & \textbf{What the phase adds} \\
\midrule
admission & caller identity, execution kind, requested resources, optional scope and governance hints & first stable envelope identity and request-time description \\
resolved & all admission fields unchanged & granted resources, routing method, backend-resolved facts \\
finalized & all earlier fields unchanged & concrete deployment identity, serve path, public path, terminal success status \\
failed & all admission fields unchanged & explicit failure stage and reason without inventing granted resources that never existed \\
\bottomrule
\end{tabular}
\end{table}

\subsection{Admission}

Admission is the only phase that should speak for caller intent. The envelope builder copies authenticated requester context, execution shape, optional scope, governance hints, and the raw resource ask into one object. At this point the backend has not yet validated placement feasibility or chosen a runtime target. That separation is deliberate. It ensures that later narrowing never rewrites the original request.

\subsection{Resolved}

Resolved is the first backend-truth phase. In the proving-ground path, serving logic enriches the envelope with granted GPU allocation, resolved engine details, and routing facts after validation. This is where the requested-versus-granted split becomes operationally useful. Operators can see not only that a deployment request eventually succeeded, but that the backend accepted a narrower or more concrete execution plan than the caller originally described.

\subsection{Finalized and failed}

Finalized is the first phase where the request becomes attached to a concrete runtime identity. Deployment identifier, serve path, public path, and final backend kind all belong here rather than in the original admission object. Failed is equally important. Many systems log failure only as a disconnected error string. The execution envelope gives failed requests a durable identity and preserves the original caller ask even when no deployment identity is ever produced. That distinction matters if later operators need to tell apart ``never granted'' from ``granted and later unhealthy.''

\section{Why \texttt{deploy\_model} First}

The first proving ground should be a path with clear execution semantics and a real resource ask. In the motivating backend, \texttt{POST /serving/deploy\_model} is the strongest candidate for several reasons:

\begin{itemize}
    \item it already exposes the richest request-time resource fields;
    \item it sits at a clean backend admission seam;
    \item it can adopt the envelope without changing placement behavior;
    \item it already needs better resource accounting and internal observability.
\end{itemize}

This is not a claim that deployment admission is the most important long-term execution path. It is a claim about implementation strategy. Deployment requests already carry explicit resource asks, already terminate in consequential backend allocation, and already need better auditability around requested-versus-granted drift. That makes them a safer proving ground than highly dynamic inference or tool-execution paths where the execution boundary itself is less crisp. In other words, \texttt{deploy\_model} is not the whole story. It is the best first slice because it forces the contract to survive a real backend resolution path.

\section{Relationship to Adjacent Systems}

\subsection{Authorization and usage control}

Usage-control work such as UCON adds continuity, mutability, and obligation semantics to access control \cite{park2004ucon}. Modern authorization languages such as Cedar emphasize analyzability and speed for complex policy decisions \cite{cutler2024cedar}. Large production authorization systems such as Zanzibar show how identity-to-object authorization can be made globally consistent at scale \cite{pang2019zanzibar}. The execution envelope complements rather than replaces these systems. Authorization decides whether a principal is allowed to request an action. The envelope describes the request in a normalized way so that such decisions, and later observability, can attach to the same object.

\subsection{Admission control}

Container orchestration systems such as Kubernetes distinguish admission from scheduling in their control plane architecture \cite{kubernetes2026admission}. That distinction is useful here. The execution envelope lives at admission time. It exists to capture normalized request semantics before backend-specific scheduling, routing, or binding decisions occur.

\subsection{Scheduling}

Large cluster schedulers such as Omega and Borg solve different problems \cite{schwarzkopf2013omega,verma2015borg}. More recent LLM-serving systems extend that line with admission-aware queueing and SLO-oriented scheduling \cite{wu2025scorpio}. They decide where and when work should run under shared resource constraints. The execution envelope explicitly avoids claiming that role. Its ambition is narrower: preserve a consistent contract describing the request and its eventual resolution so that scheduling, logging, and governance systems do not each reconstruct the same meaning independently.

\subsection{Adjacent governance substrate work}

The execution envelope also sits near a cluster of governance problems that are not identical to admission normalization but depend on it. Authorization propagation work asks how identity and authority should survive multi-agent delegation \cite{tallam2026authprop}. Fail-and-report asks how systems should surface incomplete outcomes when policy boundaries or missing authority materially affect what was produced \cite{tallam2026failandreport}. Partial Evidence Bench measures one concrete failure mode where limited evidence can still yield answers that look complete \cite{tallam2026partialbench}. Identity-conditioned delegation work asks not only what an agent \emph{can} do, but what it is appropriate for that agent to do on behalf of a particular requester \cite{tallam2026fromcantowould}. The execution envelope does not solve those problems, but it provides a seam where many of them can attach more cleanly.

\section{Evaluation Criteria}

Because the proposal is a systems primitive rather than a learning algorithm, its evaluation should focus on operational properties.

\begin{table}[t]
\centering
\small
\caption{Evaluation criteria for the execution envelope.}
\label{tab:criteria}
\begin{tabular}{>{\raggedright\arraybackslash}p{3.0cm} >{\raggedright\arraybackslash}p{10.0cm}}
\toprule
\textbf{Criterion} & \textbf{Question} \\
\midrule
Coverage & How many execution-capable backend paths can build a valid envelope without schema drift? \\
Fidelity & Does the envelope preserve both caller intent and backend grant without ambiguity? \\
Additivity & Can the first integration land without changing placement behavior or public API semantics? \\
Utility & Do logging, accounting, and downstream policy consumers actually use the envelope rather than bypass it? \\
Extensibility & Can later systems attach governance hints or runtime metadata without overloading the core contract? \\
\bottomrule
\end{tabular}
\end{table}

For an initial proving-ground implementation, three tests are especially important:

\begin{enumerate}
    \item envelope construction from existing request context and request payload;
    \item preservation of \texttt{resources\_requested} when a backend populates \texttt{resources\_granted};
    \item proof that adopting the envelope in \texttt{deploy\_model} does not change current placement behavior.
\end{enumerate}

Those tests are intentionally modest. A paper like this should not pretend that one successful integration proves platform-wide sufficiency. It should, however, prove that the contract is precise enough to survive one consequential backend path without hidden rewrites, schema drift, or silent coupling to a particular runtime implementation.

\section{Risks}

\subsection{The envelope becomes an unused bag of fields}

This is the most obvious risk. If no downstream consumer relies on the contract, then the envelope is just extra ceremony. The mitigation is architectural, not rhetorical: land it in one real path with one real use case such as request logging or resource accounting.

\subsection{Placement logic leaks into the contract}

If backend-specific routing or scheduling logic migrates into the envelope too early, the shared object becomes harder to reuse and more politically loaded. The requested-versus-granted split helps contain this by making backend resolution explicit rather than implicit.

\subsection{The contract gets narrowed around one product surface}

If the envelope is defined in terms of one collaboration primitive or one product surface, it will not scale to other execution paths. Optional \texttt{scope} fields mitigate this by making workspace, session, or job context attachable but not mandatory.

\subsection{A descriptive seam gets mistaken for an authority model}

The envelope records what execution was requested and what the backend granted. That makes it tempting to treat the object as if it were already an authority carrier or a complete policy substrate. It is not. If teams confuse descriptive normalization with actual permission semantics, they will over-trust the envelope and under-specify the systems that still need to decide what is allowed.

\section{Discussion}

The execution envelope sits in an unglamorous but important design space. It does not promise autonomous orchestration, intelligent placement, or universal governance. Instead, it tries to solve a smaller and more common systems failure: important backend concerns appear only after the original execution request has already fragmented into service-specific code paths.

That is why the paper's strongest claim is narrow. A shared admission contract is useful because it establishes one descriptive seam before those fragments occur. Once that seam exists, later systems can reason more honestly about who asked for what, under what scope, and how the backend actually resolved the request.

This also clarifies what should remain separate. An authority token that monotonically narrows permissions across execution hops is plausibly valuable, but it is answering a different question. Likewise, a scheduler or placement engine is important, but it should consume the contract rather than be embedded in it. The execution envelope is the connective tissue, not the final policy or placement brain.

\subsection{What current platforms need}

For present-day agentic platforms, the practical need is not an elegant theory of admission. It is a durable operational seam. At minimum, platforms need one place where caller identity, requested execution, optional scope, governance hints, original resource ask, backend grant, and runtime resolution can remain connected. Without that seam, every later concern---billing, audit, guardrails, postmortem review, and eventually delegated authority checks---starts by reconstructing a request the platform already processed once.

The execution envelope is useful precisely because it lets platforms add that seam without pretending to finish the rest of the stack. A platform can still have imperfect scheduling, incomplete governance policy, or weak downstream observability and benefit from the contract. The gain is that later consumers no longer have to infer the request from inconsistent service-local artifacts.

\subsection{What this paper does not claim}

This paper does not claim that admission normalization is sufficient for backend governance. It is not. It does not remove the need for explicit authority propagation, stable runtime identity, or clearer degraded-path semantics when execution is narrowed or denied. It also does not claim that one deployment path generalizes automatically to inference, tool execution, or data movement. The paper's stronger and more defensible claim is that those later problems are easier to solve honestly once there is a common object describing what execution was admitted in the first place.

\section{Future Work}

Several next steps would make the primitive materially stronger.

\textbf{Broader execution coverage.} The most obvious extension is beyond deployment admission into inference, pipeline, tool, and data-movement paths. Those paths have weaker or more dynamic resource asks, which makes them better second consumers than first ones.

\textbf{Authority-carrying attachment points.} The execution envelope is descriptive. Later work should test how it can carry references to stronger authority artifacts rather than being mistaken for one itself. That includes delegated-identity and propagation work \cite{tallam2026authprop,tallam2026fromcantowould}.

\textbf{Failure and incompleteness handling.} The proving-ground implementation already distinguishes success-path and failed-path envelopes. A natural next step is to connect the contract to more explicit degraded-path reporting so that narrowed, denied, or incomplete execution outcomes become first-class operator-visible artifacts \cite{tallam2026failandreport}.

\textbf{Evaluation surfaces.} The envelope will become more compelling when downstream systems consume it for measurable behaviors: resource accounting, request logging, postmortem traceability, and incomplete-evidence handling. Benchmark and evaluation work should test whether the contract actually reduces reconstruction overhead and unsafe interpretive drift \cite{tallam2026partialbench}.

\section{Conclusion}

AI backends are under pressure to add governance, observability, and resource-awareness at admission time, but many still lack a shared contract for the execution request itself. The execution envelope is a narrow proposal for that gap: one internal object that records who asked for what kind of execution, what resources were requested, and what the backend actually granted.

The proposal is valuable precisely because it is limited. It does not attempt to unify scheduling, authorization, and execution policy in one leap. It offers a common descriptive primitive that those systems can later consume. If the contract is right, then one proving-ground integration can make later logging, accounting, and governance work much easier. If it is wrong, it can be falsified early on a single backend path without forcing a platform-wide migration.

That is the right level of ambition for the near-term problem. Before an AI platform tries to govern all execution, it should at least be able to describe execution consistently.

\appendix

\section{Minimal artifact files}

The repository includes two concrete artifact files intended to make the proposal less purely rhetorical:

\begin{itemize}
    \item \path{execution_envelope_example.json}: an illustrative \texttt{deploy\_model} envelope instance with concrete requested and granted values;
    \item \path{execution_envelope_schema.json}: a minimal JSON Schema sketch for the shared envelope contract;
    \item \path{examples/}: resolved, finalized, and failed lifecycle examples for the same request identity.
\end{itemize}

These files are not a substitute for implementation in a live backend path, but they do force the paper's field semantics into a typed artifact that can be inspected, validated, and revised.

\bibliographystyle{plainnat}
\bibliography{references}

@article{park2004ucon,
  author = {Park, Jaehong and Sandhu, Ravi},
  title = {The UCON ABC Usage Control Model},
  journal = {ACM Transactions on Information and System Security},
  volume = {7},
  number = {1},
  pages = {128--174},
  year = {2004}
}

@article{cutler2024cedar,
  author = {Cutler, Joseph W. and Disselkoen, Craig and Eline, Aaron and He, Shaobo and Headley, Kyle and Hicks, Michael and Hietala, Kesha and Ioannidis, Eleftherios and Kastner, John and Mamat, Anwar and others},
  title = {Cedar: A New Language for Expressive, Fast, Safe, and Analyzable Authorization},
  journal = {arXiv preprint arXiv:2403.04651},
  year = {2024}
}

@inproceedings{pang2019zanzibar,
  author = {Pang, Ruoming and Allwein, Greg and Arsene, Victor and Attiyah, Kenny and Beauchamp, Robert and Bocanegra, Saulo and others},
  title = {Zanzibar: Google's Consistent, Global Authorization System},
  booktitle = {Proceedings of the 2019 USENIX Annual Technical Conference},
  pages = {33--46},
  year = {2019}
}

@misc{kubernetes2026admission,
  author = {{Kubernetes Authors}},
  title = {Admission Controllers},
  year = {2026},
  howpublished = {\url{https://kubernetes.io/docs/reference/access-authn-authz/admission-controllers/}},
  note = {Kubernetes documentation}
}

@inproceedings{schwarzkopf2013omega,
  author = {Schwarzkopf, Malte and Konwinski, Andy and Abd-El-Malek, Michael and Wilkes, John},
  title = {Omega: Flexible, Scalable Schedulers for Large Compute Clusters},
  booktitle = {Proceedings of the 8th ACM European Conference on Computer Systems},
  year = {2013}
}

@inproceedings{verma2015borg,
  author = {Verma, Abhishek and Pedrosa, Luis and Korupolu, Madhukar and Oppenheimer, David and Tune, Eric and Wilkes, John},
  title = {Large-Scale Cluster Management at Google with Borg},
  booktitle = {Proceedings of the Tenth European Conference on Computer Systems},
  year = {2015}
}

@article{wu2025scorpio,
  author = {Tang, Yinghao and Lan, Tingfeng and Huang, Xiuqi and Lu, Hui and Chen, Wei},
  title = {SCORPIO: Serving the Right Requests at the Right Time for Heterogeneous SLOs in LLM Inference},
  journal = {arXiv preprint arXiv:2505.23022},
  year = {2025}
}

@misc{tallam2026authprop,
  author = {Tallam, Krti},
  title = {Authorization Propagation in Multi-Agent AI Systems: Identity Governance as Infrastructure},
  year = {2026},
  note = {Unpublished manuscript}
}

@misc{tallam2026failandreport,
  author = {Tallam, Krti},
  title = {Fail-and-Report: A Missing Authorization Primitive for Agentic AI Systems},
  year = {2026},
  note = {Unpublished manuscript}
}

@misc{tallam2026partialbench,
  author = {Tallam, Krti},
  title = {Partial Evidence Bench: Benchmarking Authorization-Limited Evidence in Agentic Systems},
  year = {2026},
  note = {Unpublished manuscript}
}

@misc{tallam2026fromcantowould,
  author = {Tallam, Krti},
  title = {From Can to Would: Identity-Conditioned Authorization for Delegated Agentic Action},
  year = {2026},
  note = {Unpublished manuscript}
}

\end{document}